\documentclass[letterpaper,12pt]{article}
\usepackage{amsmath}
\usepackage{thmtools}
\usepackage{ctable,multirow}
\usepackage{subfigure}
\usepackage{threeparttable}
\usepackage[margin=10pt,labelfont=bf]{caption}
\usepackage[margin=1in]{geometry}
\usepackage{setspace}
\usepackage{parskip}
\usepackage[round]{natbib} %for bibliography

\newcommand{\bm}{\boldsymbol}
\newcommand{\bfbeta}{\mbox{\boldmath $\beta$}}
\newcommand{\bfgamma}{\mbox{\boldmath $\gamma$}}

\newcommand{\bftheta}{\mbox{\boldmath $\theta$}}
\usepackage{titlesec}
\titleformat{\section}
  {\normalfont\normalsize\bfseries}{\thesection.}{1em}{}
\titlespacing*{\section}{0pt}{*4}{*1}
\titleformat{\subsection}
  {\normalfont\normalsize\itshape}{\normalfont\thesubsection.}{1em}{}
\titlespacing*{\subsection}{0pt}{*2}{*0}
\titleformat{\subsubsection}
  {\normalfont\normalsize\itshape}{\normalfont\thesubsubsection.}{1em}{}
\titlespacing*{\subsubsection}{0pt}{*2}{*0}

\usepackage[labelsep=endash]{caption}

\date{}
\begin{document}

\begin{center}
%\title{Landmark Proportional Subdistribution Hazards Models for Dynamic Prediction of Cumulative Incidence Functions Supplementary Materials} 
{\Large \textbf{Landmark Proportional Subdistribution Hazards Models for Dynamic Prediction of Cumulative Incidence Functions}}\\

\hspace{10pt}

{\normalsize Qing Liu$^1$, Gong Tang$^{2,3}$, Joseph P. Costantino$^{2,3}$, Chung-Chou H. Chang $^{\ast2,4}$}\\

\hspace{5pt}

\small  
$^1$\textit{Amgen Inc., 1 Amgen Center Dr, Thousand Oaks, CA 91320, USA}\\
$^2$ \textit{Department of Biostatistics, University of Pittsburgh, 130 De Soto Street, Pittsburgh, PA 15261, USA}\\
$^3$ \textit{NRG Oncology Statistics and Data Management Center, 201 N Craig Street, Pittsburgh, PA 15213, USA}\\
$^4$ \textit{Department of Medicine, University of Pittsburgh, 3550 Terrace Street, Pittsburgh, PA 15261, USA}\\
$^\ast$ {changj@pitt.edu}

\end{center}

\begin{abstract}
{An individualized risk prediction model that dynamically updates the probability of a clinical event from a specific cause is valuable for physicians to be able to optimize personalized treatment strategies in real-time by incorporating all available information collected over the follow-up. However, this is more complex and challenging when competing risks are present, because it requires simultaneously updating the overall survival and the cumulative incidence functions (CIFs) while adjusting for the time-dependent covariates and time-varying covariate effects. In this study, we developed a landmark proportional subdistribution hazards (PSH) model and a more comprehensive supermodel by extending the landmark method to the Fine-Gray model. The performance of our models was assessed via simulations and through analysis of data from a multicenter clinical trial for breast cancer patients. Our proposed models have appealing advantages over other dynamic prediction models for data with competing risks. First, our models are robust against violations of the PSH assumption and can directly predict the conditional CIFs bypassing the estimation of overall survival and greatly simplify the prediction procedure. Second, our landmark PSH supermodel enables researchers to make predictions at a set of landmark points in one step. Third, the proposed models can easily incorporate various types of time-dependent information using existing standard software without computational burden. }
\end{abstract} 


\section{Introduction \label{sec:introduction}}

An individualized risk prediction model that estimates the probability of a clinical event dynamically using time-dependent information can enable physicians to optimize personalized treatment strategies in real-time. For example, for an early-stage breast cancer patient who received surgery, the physicians can adjust the approach to the subsequent chemotherapies if knowing the risk of a distant metastasis within the next few years given the timing of the patient's occurrence of locoregional recurrence. Note that the occurrence of locoregional recurrence is a time-dependent intermediate event. Other types of time-dependent information include longitudinal measurements of biomarkers and time-varying covariate effects, which are also crucial to the success of a risk prediction model. 

Two popular approaches in dynamic prediction that received significant attention are joint modeling (Proust-Lima and Taylor 2009, Rizopoulos 2011, Mauguen et al.\ 2013, Rizopoulos et al.\ 2014) and landmarking (van Houwelingen 2007, van Houwelingen and Putter 2008 \& 2012, Parast et al.\ 2012). To incorporate the time-dependent information, joint modeling captures the processes of longitudinal and time-to-event data simultaneously whereas the landmark method applies time-to-event modeling on patients who are still at risk at the updated prediction baseline, i.e., the landmark time, using the information collected up to the landmark time. Although the joint models provide additional information from the joint distributions, the implementation is complex and the correlation structure is difficult to specify correctly thus often results in inaccurate prediction. By contrast, the landmark Cox model and supermodels proposed by van Houwelingen (2007) showed significant advantages in simplicity, flexibility, and accuracy in estimating the risk. Moreover, they become robust against model misspecifications by avoiding the full specification of the joint distribution. These models can easily incorporate various types of time-dependent information without using computationally intensive estimation procedures, and the estimation can be implemented in everyday clinical practice by using standard statistical software. Recently, landmark modeling was used in many clinical applications (Zamboni et al.\ 2010, Fontein et al.\ 2015). 

To account for data with competing events, Cortese and Andersen (2009) and Nicolaie et al.\ (2013a) applied the landmark method onto the cause-specific hazards model. The disadvantage of this approach is that to estimate the risk from one cause we need to construct the landmark models or supermodels based on the cause-specific hazards for all causes. For direct modeling cause-specific risk, the Fine-Gray model (Fine and Gray 1999), pseudo-observation approach (Klein and Andersen 2005), and direct binomial regression model (Scheike et al.\ 2008) are the commonly used methods, and the Fine-Gray model is the most widely used one because the estimated risk can be easily calculated. Cortese et al.\ (2013) applied the landmark method to the Fine-Gray model at a small set of fixed landmark time points, however, their method did not assess robustness of their proposed model against violations of the proportional subdistribution hazards (PSH) assumption, nor did they construct the comprehensive landmark supermodel to estimate the risks of a sequence of prediction time points simultaneously. The major challenge encountered in developing a landmark model/supermodel based on the Fine-Gray model was in constructing the landmark subset at each landmark time to properly account for competing events in the setting of subdistribution hazard (Nicolaie et al.\ 2013a). Nicolaie et al.\ (2013b) developed a landmark supermodel based on the pseudo-observations created at each landmark time point, however, additional steps are required to calculate the risks. 

In this study, we overcame the aforementioned challenges, extended the landmark method to the Fine-Gray model, and proposed landmark PSH model and landmark PSH supermodel for dynamic prediction in the presence of competing risks. Our proposed models can directly predict the risk for a specific cause and provide accurate prediction even when the PSH assumption fails to hold. The landmark PSH supermodel uses smoothing techniques to aggregate a set of simple landmark PSH models, allowing users to make dynamic predictions at arbitrary landmark points using only one model, also provides a simple and explicit estimation form to incorporate time-dependent covariates and/or time-varying covariate effects. Compared to the currently available landmark dynamic prediction models for competing risks, our proposed models are more straightforward and can be effortlessly implemented with standard statistical software. 

This paper is organized as follows. In Section~\ref{sec: dynpred}, we first introduce notations and dynamic predictive quantities for data with competing risks,  and then review the Fine-Gray PSH model and present our proposed landmark PSH model and supermodel. In Section~\ref{sec: simulations}, we assess the performance in prediction for our proposed models and compare with the existing methods using simulations. In Section~\ref{sec: application}, we apply the proposed models to predict the dynamic cumulative incidences of distant metastasis based on the locoregional recurrence status and a set of prognostic factors of breast cancer patients in a multicenter clinical trial. The discussions are provided in Section~\ref{sec: discussion}.

%******************************************
%    Notations and Conditional CIF
%******************************************

\section{Dynamic Prediction Models \label{sec: dynpred}}

\subsection{Notations and Conditional CIF}
\label{sec: notations}
Let $T$ and $C$ be the failure and censoring time, respectively; $\varepsilon\in \{1, \ldots , k\}$ be the cause of failure; and $\bm{Z}(\cdot)$ be a $p$-dimensional vector of covariates, which could be time-fixed covariates measured at the baseline or be time-dependent covariates. Here, for each subject $i$ $(i=1, \ldots , n)$, we assume $C_{i}$ is independent of $T_{i}$ and $\bm{Z}_{i}(\cdot)$, and refer to it as random censoring. For right censored data, we observe an independently and identically distributed quadruplet of $\{X_{i}=T_{i}\wedge C_{i}, \Delta_{i}=I(T_{i}\leq C_{i}), \Delta_{i}\varepsilon_{i}, \bm{Z}_{i}(\cdot)\}$. Suppose cause $1$ is the primary event of interest, the subdistribution function or CIF of cause $1$ is defined as $F_{1}(t)=\Pr(T\leq t,\varepsilon=1)$. 

In dynamic prediction, the aim is to predict the conditional CIF that is the cumulative incidence of occurring the primary event within the next $w$ time units (also named as a prediction window with a fixed width of $w$) for a subject who has not failed from any cause at a landmark time (denoted by $s$) by incorporating the time-dependent information available up to the landmark time $s$. The width of prediction window $w$ depends on the clinical relevance that defines the time interval within which the risk of having the primary event is of interest. Then, the conditional CIF is defined as
\begin{eqnarray}
      F_{1,LM}\{s+w|\bm{Z}(s), s\} = \Pr\{T\leq s+w, \varepsilon=1|T > s,\bm{Z}(s)\}, \nonumber
%      \label{eq: conditionalCIF}
\end{eqnarray}
where $\bm{Z}(s)$ is the covariates $\bm{Z}(\cdot)$ whose values are measured or available at $s$. Using the definition of conditional probabilities, we can rewrite the conditional CIF as 
\begin{eqnarray}
      F_{1,LM}\{s+w|\bm{Z}(s), s\} = \frac{F_{1}\{s+w| \bm{Z}(s)\} - F_{1}\{s| \bm{Z}(s)\}}{S\{s| \bm{Z}(s)\}} = \frac{F_{1}\{s+w| \bm{Z}(s)\} - F_{1}\{s| \bm{Z}(s)\}}{1 - \sum_{j=1}^{k}F_{j}\{s| \bm{Z}(s)\}}, \nonumber
\end{eqnarray}
where $F_{j}\{s| \bm{Z}(s)\}$ is the CIF for cause $j$ and $S\{s| \bm{Z}(s)\}$ is the overall survival at time $s$.

Though existing methods can be used to estimate $ F_{1,LM}\{s+w|\bm{Z}(s), s\}$, the disadvantage is obvious. Regression models (e.g. cause-specific hazards model and the Fine-Gray model) are required to estimate the CIFs for all causes. In order to directly estimate the conditional CIF without fitting models for all causes, we extend the landmark method developed by van Houwelingen (2007) to the competing risks setting and propose a landmark PSH model which is also robust against the misspecification of the proportional subdistribution hazards.

%******************************************
%    Dynamic Predictive Models
%******************************************

\subsection{Fine-Gray Model}
Fine and Gray (1999) proposed a regression model with respect to the subdistribution hazard $\lambda_1(t) = \lim_{\Delta t \to 0}\frac{1}{\Delta t} \Pr \{t \leq T \leq t+\Delta t, \varepsilon = 1|T \geq t \cup (T \leq t \cap\varepsilon \ne 1 )\} $. %which is defined by Gray (1988). The Fine-Gray model takes the form $\lambda_1(t|\bm{Z}) =\lambda_{10}(t) \exp( \bm{Z}^T\bfbeta)$ where $\lambda_{10}(t)=d \log\{1-F_{1}(t|\bm{Z}=\underline{0})\}/ dt$ is an unspecified, nonnegative function for the baseline subdistribution hazard, $\bm{Z}$ is a vector of time-independent covariates, and where $\bfbeta$ is a $p\times 1$ vector of unknown regression parameters. The CIF of cause $1$, $F_1(t|\bm{Z})$, can then be calculated as $ 1-\exp\{-\int_0^t \lambda_{10}(u) \exp( \bm{Z}^T\bfbeta) \, du \}$. 
With time-dependent covariates $\bm{Z}(t)$, the Fine-Gray model takes the form $\lambda_1\{t|\bm{Z}(t)\} =\lambda_{10}(t) \exp\{ \bm{Z}(t)^T\bfbeta\}$; and $F_1\{t|\bm{Z}(t)\} = 1-\exp\{-\int_0^t \lambda_{10}(u) \exp\{ \bm{Z}(u)^T\bfbeta\} \, du \}$. The regression coefficients $\bfbeta$ are estimated through a partial likelihood approach with modified risk sets which are defined as $R(T_{i})=\{j: (T_{j}\geq T_{i}) \cup (T_{j} \leq T_{i} \cap \varepsilon_{j} \ne 1)\}$ for the $i$th individual. $R(T_{i})$ includes all subjects who have not failed from the event of interest by time $T_{i}$. %When random right censoring is present,  the inverse probability of censoring weighting (IPCW) technique (Robins and Rotnitzky 1992) is applied to obtain an unbiased partial likelihood estimator $\hat{\bfbeta}_{PL}$ via a weighted score function, where the weight is defined as $\omega_i(t)= I(C_i \geq T_i \wedge t) \hat{G}(t)/\hat{G}(X_i \wedge t)$ for the $i$th individual and $\hat{G}(t)$ is the Kaplan-Meier estimator of the censoring survival distribution $G(t)=\Pr(C\geq t)$. Then $F_1(t|\bm{Z})$ can be estimated by calculating $\hat{F}_1(t|\bm{Z})=1-\exp\left\{-\exp(\bm{Z}^T\hat{\bfbeta}_{PL}) \hat{\Lambda}_{10}(t)\right\}$, where $\hat{\Lambda}_{10}(t)$ is a modified version of the Breslow estimator. %It has been shown that $\hat{F}_1(t;\bZ)$ uniformly converges to $F_1(t;\bZ)$ on the interval $[0,\tau)$, where $\tau=\sup\{t:\Pr(X\geq t)>0\}$ \citep{fine1999proportional,Geskus2011}. 

\subsection{Landmark PSH Model}
To directly estimate the conditional CIF $ F_{1,LM}\{s+w|\bm{Z}(s), s\}$ at a given landmark time $s$, we define a conditional subdistribution hazard
\begin{equation}
\lambda_1\{t|\bm{Z}(s),s\} = \lim_{\Delta t \to 0}\frac{1}{\Delta t} \Pr \{t \leq T \leq t+\Delta t, \varepsilon = 1|T \geq t \cup (s \leq T \leq t \cap\varepsilon \ne 1 ), \bm{Z}(s)\} 
\label{eq:consubhaz}
\end{equation}
for all $t \geq s$, which is the subdistribution hazard $\lambda_{1}\{t|\bm{Z}(s)\}$ conditional on assuming no event occurred from any cause prior to $s$. It can be shown that (\ref{eq:consubhaz}) is the hazard function of the conditional CIF:
\begin{eqnarray}
	F_{1,LM}\{s+w|\bm{Z}(s), s\} &=& \Pr\{s < T \leq s+w, \varepsilon = 1|T > s,\bm{Z}(s)\} \nonumber \\
	&=& 1 - \Pr\{(T > s+w )\cup (s < T \leq s+w \cap \varepsilon \ne 1)|T > s,\bm{Z}(s)\} \nonumber \\
    &=& 1 - \exp (-[\Lambda_{1}\{s+w|\bm{Z}(s), s\}-\Lambda_{1}\{s-|\bm{Z}(s), s\}]) \nonumber \\
	&=&  1 - \exp\left[ -\int_{s}^{s+w} \lambda_1\{t|\bm{Z}(s), s\} \, dt\right], \nonumber
\end{eqnarray}
where $\Lambda_{1}\{\cdot|\bm{Z}(s), s\}$ is a cumulative conditional subdistribution hazard at landmark time $s$. Therefore, we propose a \textit{landmark PSH model} with the form
\begin{eqnarray}
   \lambda_1\{t|\bm{Z}(s), s\}&=& \lambda_{10}(t|s) \exp\{\bm{Z}(s)^T \bfbeta_{LM}\},
  \label{eq:crudeLMFG}
\end{eqnarray}
for $s \leq t \leq s+w$. 

To estimate the regression parameters $\bfbeta_{LM}$ and the baseline conditional subdistribution hazard $\lambda_{10}(t|s)$, we can simply apply the Fine-Gray model to subjects who have not yet failed from any cause at $s$ and ignoring the events occurring after $s+w$ by adding an administrative censoring at $s+w$. As discussed in Liu et al. (2016), under non-PSH, the Fine-Gray model with an administrative censoring at the horizon time provides an accurate prediction of CIF at this horizon time (but not other time points) without constructing complex procedures to estimate time-varying effects; and the partial likelihood estimator $\hat{\bfbeta}_{LM}$ of $\bfbeta_{LM}$ in model (\ref{eq:crudeLMFG}) is consistent for a weighted average of possibly time-varying effect over the interval $[s, s+w]$. Thus, the proposed landmark PSH model with truncation at the landmark time $s$ and administrative censoring at the horizon time $s+w$ is robust to violations of the PSH assumption. The conditional CIF can be easily estimated as
\begin{eqnarray}
\hat{F}_{1,LM}\{s+w|\bm{Z}(s), s\}=1-\exp\left[-\exp\{\bm{Z}(s)^T\hat{\bfbeta}_{LM}\} \{\hat{\Lambda}_{10}(s+w)-\hat{\Lambda}_{10}(s-)\}\right]\nonumber, 
\end{eqnarray}
in which $\hat{\bfbeta}_{LM}$ and $\hat{\Lambda}_{10}(\cdot)$ are estimates calculated from the Fine-Gray model. Therefore, the landmark PSH model provides a simple and convenient way to directly predict the conditional CIF in dynamic prediction.

\subsection{Landmark PSH Supermodel \label{subsec: lmfgsuper}}
For dynamic prediction, it is required to estimate conditional CIFs at several different landmark time points because the intermediate events may occur at any time point and covariate values may also be updated during the follow up. Therefore, it is of interest to dynamically predict the risk probabilities $F_{1,LM}\{s+w|\bm{Z}(s), s\}$ with fixed width $w$ for varying landmark time $s$ within an interval $[s_{0}, s_{L}]$. As discussed in van Houwelingen and Putter (2012), this provides a dynamic prediction framework with a sliding prediction window over the specified interval. Theoretically we can fit $L+1$ landmark PSH models (\ref{eq:crudeLMFG}) to obtain $L+1$ conditional CIFs. When the number of landmark points increased dramatically, it is less practical and computationally inefficient to fit many models separately.  

Instead, we adopt the smoothing strategy of van Houwelingen (2007) and propose a \textit{landmark PSH supermodel} with the form
 \begin{equation}
    \lambda_1\{t|\bm{Z}(s),s\} =\lambda_{10}(t) \exp\{\bm{Z}(s)^T \bfbeta_{LM}(s)+\bfgamma(s)\}
   \label{eq:pbLMFG}
 \end{equation}
 for $s_{0} \leq t \leq s_{L}+w$, which can estimate conditional CIFs at a set of landmark time points from $[s_{0}, s_{L}]$ in one step. As shown in Liu et al.\ (2016),  $\hat{\bfbeta}_{LM}$ of $\bfbeta_{LM}$ in model (\ref{eq:crudeLMFG}) is consistent for a weighted average of $\bfbeta(t)$ over the interval $[s, s+w]$; so we can expect the effect of $s$ on $\bfbeta_{LM}(s)$ to be a continuous function and model $\bfbeta_{LM}(s)$ as continuous functions of $s$. For $\lambda_{10}(t|s)$, as the Breslow estimator showed that the dependence on $s$ is through $\bfbeta_{LM}(s)$, we can also expect that the $\lambda_{10}(t|s)$ varies continuously with $s$ and rewrite $\lambda_{10}(t|s)$ as $\lambda_{10}(t|s) = \lambda_{10}(t)\exp\{\bfgamma(s)\}$. Hence, the sequence of landmark PSH models at a set of landmark time points are combined into a supermodel via $\bfbeta_{LM}(s)$ and $\bfgamma(s)$ which are both continuous functions of $s$. For simplicity, we can fit linear models for $\bfbeta_{LM}(s)$ and $\bfgamma(s)$ as following forms $\bfbeta_{LM}(s) = \bfbeta_{LM}(s|\bftheta) = \sum_{j=1}^{m_{\beta}} \bftheta_{j}f_{j}(s)$ and $\bfgamma(s) = \bfgamma(s|\bm{\eta}) = \sum_{j=1}^{m_{\lambda_{10}}} \eta_{j}g_{j}(s), $ 
where $f(s)$ and $g(s)$ are two sets of parametric functions of $s$, $\bftheta$ and $\bm{\eta}$ are vectors of parameters, and $m_{\beta}$ and $m_{\lambda_{10}}$ are the number of parameters in $\bftheta$ and $\bm{\eta}$.  Note that $g(s)$ do not need to contain the constant term and have the restriction of $g_{j}(s_{0})=0$ for all $j$ $(j = 1, \dots, m_{\lambda_{10}})$  due to identifiability of the baseline subdistribution hazard, as discussed in van Houwelingen (2007).

To fit the landmark PSH supermodel, similar to the landmark supermodel for survival data without competing risks (van Houwelingen 2007),  we need to create an augmented dataset which is constructed as follows: first select a set of landmark points $s$ from the interval $[s_{0}, s_{L}]$; for each $s$, create a landmark subset by selecting the subjects who have not yet failed from any cause at $s$ and adding an administrative censoring at the prediction horizon $s+w$; and then stacking all the individual landmarking subsets into a super prediction dataset. Then, model (\ref{eq:pbLMFG}) can be fitted by applying a Fine-Gray model including landmark-covariate interactions $\bm{Z}(s)*f_{j}(s)$ to the stacked dataset. 

The parameters $(\bftheta, \bm{\eta})$ can be consistently estimated by maximizing a Breslow pseudo partial log-likelihood for tied events, because that in the stacked dataset, one subject with event time $T_{i}$ has $ n_{i} = \#\{s: s \leq T_{i} \leq s+w, s \in [s_{0}, s_{L}]\}$ repeated observations; and $\hat{\Lambda}_{10}(t)$ can also be estimated by a Breslow estimator for the cumulative subdistribution hazard (see Web Appendix A for the details). 
To obtain the standard errors for the estimated parameters, a robust sandwich estimator is required to adjust for the correlation between the risk sets which exists because the same subject is repeatedly used when we estimate the parameters based on the stacked dataset. Thus, the target dynamic prediction probabilities $F_{1,LM}\{s+w|\bm{Z}(s), s\}$ have a simple and explicit estimation form, which is given by
\begin{equation}
 \hat{F}_{1,LM}\{s+w|\bm{Z}(s), s\} = 1 - \exp\left[-e^{\bm{Z}(s)^{T}\hat{\bfbeta}_{LM}(s)+\hat{\bfgamma}(s)}\{\hat{\Lambda}_{10}(s+w)-\hat{\Lambda}_{10}(s-)\}\right]
\label{eq: predcondCIF}
\end{equation}
for all $s \in [s_{0}, s_{L}]$. Therefore, the landmark PSH supermodel can provide the prediction of $F_{1,LM}\{s+w|\bm{Z}(s), s\}$ in any period of length $w$ starting anywhere in the interval $[s_{0}, s_{L}]$. 

%Note that model \eqref{eq:pbLMFG} assume that the effect of $s$ on $\lambda_{10}(t|s, w)$ in an additive way. As discussed in \citet{vanHou2007}, this assumption hold if the follow-up is not too long or the effect of covariates is not too big. So, if we choose an optimal width $w$ for the prediction window and a rational range $[s_{0}, s_{L}]$ for the landmark time points, the PBLM-PSH supermodel directly provides a correct approximation for the conditional CIF at time $s+w$ for any $s\in [s_{0}, s_{L}]$.  

For implementation, before stacking all landmarking subsets into a super dataset,  we need to transform each subset into the counting process style and include time-varying IPCW weights for the subjects experienced competing events to adjust for the random right censoring (Geskus 2011). In practice, fitting the landmark PSH supermodel in the stacked dataset requires a software that allows for delayed entry or left truncation at $s$. The \texttt{R} function \texttt{coxph()} can be used to fit model (\ref{eq:pbLMFG}) and it can also provide robust sandwich covariance matrix for $(\bftheta, \bm{\eta})$ which can be used in the significance test of the estimated regression coefficients. As discussed, the landmark effect on $\lambda_{10}(t|s)$ is through $\bfbeta_{LM}(s)$, so that there is a correlation between $\bftheta$ and $\bm{\eta}$. It is suggested to center the covariates before fitting the model (van Houwelingen and Putter 2012).  

For simplicity, we assumed that $C$ is independent of $T$ and $\bm{Z}$, but, it can be generalized to a conditional independence between $C$ and $T$ given $\bm{Z}$, as discussed in the PSH model (Fine and Gray, 1999). The dependence between $C$ and $\bm{Z}$ can be handled by modeling $C$ as a function of $\bm{Z}$ in IPCW. Following the same strategy, the properties of landmark PSH model and landmark PSH supermodel will be retained under conditional independent censoring.

\subsection{Measure of Predictive Performance}
\label{sec: tdBrierscore}
To evaluate dynamic predictive performance of the proposed procedures, we adapted the time-dependent Brier score, O/E ratio, and the area under the receiver operating characteristics curve (AUC) in terms of predictive accuracy, calibration, and discrimination, respectively.  

The time-dependent Brier score is an estimate of the mean-squared prediction errors of the predicted event probabilities at $s+w$ over the observed event status for subjects who are event free at landmark time $s$ (Schoop et al.\ 2011, Cortese et al.\ 2013). For data with competing risks, the expected time-dependent Brier score at landmark time $s$ for the prediction at $s+w$ is defined as
\begin{eqnarray}
       \text{BS}_{LM}(s,w)=E\left( [I(T\leq s+w, \varepsilon=1)-F_{1,LM}\{s+w|\bm{Z}(s),s\}]^2|T>s\right).  \nonumber
\end{eqnarray}
When random right censoring exists, we utilized a pseudovalue-based consistent estimator given by $\widehat{\text{BS}}_{LM}(s,w)={\tilde{n}_{s}}^{-1}\sum_{i \in \tilde{R}_{s}}^{}\left( \hat{Q}_{1,LM}^{(i)}(s+w|s)[1-2\hat{F}_{1,LM}\{s+w|\bm{Z}_{i}(s),s\}] + \hat{F}_{1,LM}\{s+w|\bm{Z}_{i}(s),s\}^2 \right)$, which was proposed by Cortese et al. (2013), 
%\begin{eqnarray}
%       \widehat{\bm{Brier}}_{LM}(s,w)=\frac{1}{\tilde{n}_{s}}\sum_{i \in \tilde{R}_{s}}^{}\left( \hat{Q}_{1,LM}^{(i)}(s+w|s)[1-2\hat{F}_{1,LM}\{s+w|\bm{Z}_{i}(s),s\}] + \hat{F}_{1,LM}\{s+w|\bm{Z}_{i}(s),s\}^2 \right), \nonumber
%\end{eqnarray}
where $\tilde{R}_{s}=\{i: X_{i} > s\}$ and $\tilde{n}_{s}$ is the number of subjects in $\tilde{R}_{s}$. $\hat{Q}_{1,LM}^{(i)}(s+w|s)= \tilde{n}_s\hat{F}_{1,LM}(s+w|s)-(\tilde{n}_{s}-1)\hat{F}_{1,LM}^{(i)}(s+w|s)$ is a jackknife pseudovalue for the $i$th subject who is still at risk at time $s$, where $\hat{F}_{1,LM}(s+w|s)$ is the nonparametric estimate of the marginal cumulative incidence $\Pr(T\leq s+w|T>s)$, and $\hat{F}_{1,LM}^{(i)}(s+w|s)$ is the same estimate but is based on the data where the $i$th subject has been removed. 

To assess calibration, we adapted the time-dependent O/E ratio which is defined as the observed number of main events from the landmark time point divided by the prediction horizon time to the expected number of events estimated from the predictive model,
\begin{eqnarray}
       \text{O/E}_{LM}(s,w)=\frac{\sum{I(s<T_i\leq s+w, \varepsilon=1)}}{\sum{F_{1,LM}\{s+w|\bm{Z_i}(s),s\}}}.  \nonumber
\end{eqnarray}
As discussed in Pfeiffer and Gail (2017), to address the right censoring, the jackknife pseudovalue $\hat{Q}_{1,LM}^{(i)}(s+w|s)$ was applied, and $\widehat{\text{O/E}}_{LM}(s,w) = \frac{\sum_{i \in \tilde{R}_{s}}^{}{\hat{Q}_{1,LM}^{(i)}(s+w|s)}}{\sum_{i \in \tilde{R}_{s}}^{}{\hat{F}_{1,LM}\{s+w|\bm{Z_i}(s),s\}}}.$

To measure discrimination capability of the landmark PSH supermodel, we followed the work of Blanche et al. (2015) and Huang et al. (2016) to calculate the time-dependent AUC which is given by
\begin{equation}
\bm{\text{AUC}}_{LM}(s, w) =\Pr \{ \pi_{i}(s, w) > \pi_{j}(s, w) | D_i(s,w)=1, D_j(s,w)=0, T_i > s, T_j > s\}   \nonumber
\end{equation} 
for a pair of subjects $\{i, j\}$ both of whom have not experienced any event at the landmark time $s$, where $\pi_{i}(s, w) = F_{1,LM}\{s+w|\bm{Z_i}(s),s\}$ and $D_i(s,w)=I(s < T_{i} \leq s+w, \varepsilon_{i} = 1)$. In the presence of right censoring, an inverse probability of censoring weighting (IPCW) estimator proposed by Blanche et al. (2015) was utilized.

%$(s < T_{i} \leq s+w, \varepsilon_{i} = 1) \cap \{ (s < T_{j} \leq s+w, \varepsilon_{j} \ne 1 ) \cup (T_{j} > s+w)\}] $

%******************************************
%    Simulation
%******************************************
%\clearpage
\section{Simulation Studies \label{sec: simulations}}

We evaluated the performance of the proposed dynamic prediction models using simulated data under two different settings of the nonproportional subdistribution hazards  (non-PSH).  %The time-dependent Brier scores were compared among the proposed landmark PSH model, the landmark PSH supermodel, the nonparametric method, and the standard Fine-Gray model (also named the PSH model). 
Because that nonparametric method is model-free, we used it as the reference to compare with other model-based methods in scenarios where only categorical variables are included. For simplicity, only two failure types were considered: type 1 failure is the primary event of interest; type 2 failure indicates competing events. We also simulated a competing risks data under non-PSH with a time-dependent covariate. % and calculated the AUC to assess the discriminative capability of the landmark PSH supermodel.

In the first non-PSH setting, we generated the type 1 failure times from a two-parameter Weibull mixture distribution with the subdistribution  $F_{1, 1s}(t;Z_{i})=p(1-\exp[-\{\lambda_{1} \exp(Z_{i}\beta_{1})t\}^{\alpha_1}])$, %with $(\alpha_{1}, \lambda_{1}, \beta_{1}) = (3.2, 0.18, -0.81)$
where the rate parameter depends on the covariate $Z_{i}$. In the second non-PSH setting, we let the coefficient of $Z_{i}$ be a function of time; and the subdistribution of the primary event became $F_{1, 2nd}(t;Z_{i})=1-(1-p[1-\exp\{-(\lambda_{2} t)^{\alpha_{2}}\}])^{\exp\{Z_{i}\beta_{21}+Z_{i}\beta_{22}\ln(t+1)\}}$. % with $(\alpha_{2}, \lambda_{2}, \beta_{21},\beta_{22}) = (3.2, 0.12, 0.8, 0.3)$. In both settings, we considered $Z_{i}$ as Bernoulli(0.5) variants, and let $p = 0.3$, which produced about $30 \%$ main events at $Z_{i}=0$ when there was no censoring. 
We generated the type 2 failure times from an exponential distribution $\Pr (T_{i} \leq t|\varepsilon_{i}=2, Z_{i}) = 1-\exp\{-\exp(Z_{i}\beta_{c}t)\}$ by taking $\Pr(\varepsilon_{i}=2|Z_{i}) = 1 - \Pr(\varepsilon_{i}=1|Z_{i})$, where $\beta_{c}=0.5$.  Sample size of $n=1,000$ was chosen and the data were simulated repeatedly for $N=1,000$ times. The censoring times were generated independently from a uniform distribution.

The performance of the proposed landmark PSH model and the landmark PSH supermodel in dynamic prediction of the conditional CIFs using a fixed width of $w$ at a set of landmark points were compared with the prediction performance of the nonparametric method and the standard PSH model. %We chose a prediction window of width $w=3$ for the first non-PSH setting and $w=2$ for the second non-PSH setting, 
The width of prediction window $w$ was chosen based on the distributions of the primary event of interest (Web Figure 1). 

For fitting the landmark PSH supermodel, we set up a fine grid of landmark points with equidistant step of 0.1 from 0 to 5 for the first non-PSH setting and from 0 to 4 for the second non-PSH setting. In both settings, we took ordinary polynomials for the basis functions as $f(s) = \{1, s, s^2\}$ and $g(s) = \{s, s^2\}$. Note that the landmark PSH supermodel in absence of censoring cannot be fitted using the \texttt{coxph()} function in R, because that the competing risks data cannot be transformed into the counting process format if there is no censoring; and that the model cannot be fitted by the \texttt{crr()} function in R either since \texttt{crr()} does not allow delayed entries. Therefore, in the absence of censoring, only the landmark PSH model is fitted to compare with other standard approaches. 

Figure 1 %~\ref{fig: predictedcondCIF} 
depicts the true and estimated conditional CIFs obtained from different approaches. We also provided the estimates of the conditional failure probability from the landmark Cox supermodel (van Houwelingen 2007) by treating the competing risks as random censoring. We found that in two different non-PSH scenarios, the performance of the landmark PSH model and the landmark PSH supermodel are as good as that of the nonparametric methods. But the landmark Cox supermodel overestimated the risk of failure for main event in the presence of competing risks.

%%******* Figure 0 ****************
%\begin{figure}[htbp]
%  \centering
%  \includegraphics[height=3.6in,width=3.6in]{Figure0.eps} 
%\caption{Cumulative incidence function of main event under two different non-PSH settings.}
% \label{fig: CIFsimu}
%\end{figure}

For each approach we evaluated the prediction errors in the dynamic conditional CIFs by estimating time-dependent Brier scores, and we used a 3-fold cross-validation to correct for possible overfitting. Web Table 1 presents averaged estimates of the cross-validated time-dependent Brier score and its empirical standard deviation. To quantify the improvement of predictive accuracy for the proposed landmark PSH model and supermodel to the standard PSH model under nonproportional hazards, we utilized a relative increment (or reduction) of prediction errors by treating the nonparametric estimates as the reference. The relative increment of prediction errors are presented in Figure 2. %~\ref{fig: cvbs}. 
As expected, in both non-PSH settings the predictive accuracy of the landmark PSH model was almost the same as that obtained from the nonparametric method. As compared with the landmark PSH model or the nonparametric method, the landmark PSH supermodel has slightly lower accuracy, yet the differences in prediction errors are negligible.

To further evaluate predictive performance of the proposed landmark PSH supermodel for data with time-dependent covariates, we added $Z_{i}(t)$ into the second non-PSH setting using the simulation strategy introduced in Huang et al. (2016) (see Web Appendix B for details). 
%we generated $Z_{i}(t)$ from a liner mixed effects model $Z_{i}(t) = m_{i}(t) + \epsilon_{i}(t)$, where $m_{i}(t) = (\beta_{30} + b_{i0}) + (\beta_{31} + b_{i1})t$, $(b_{i0}, b_{i1})$ have a bivariate normal distribution with mean 0 and covariance matrix  
%$\left({\begin{array}{*{20}c}
%   0.2 & 0.05  \\
%   0.05 & 0.1  \\    
% \end{array} } \right),$ the random error term $\epsilon_{i}(t)$ follows $N(0, 0.6^2)$ at each measurement time $t$, and we measured the covariate at $t = 0, 1, 2, ..., 6$. Thus, the primary event was generated from the subdistribution $F_{1, 3rd}\{t;Z_{i}, Z_{i}(t)\}=1-(1-p[1-\exp\{-(\lambda_{3} t)^{\alpha_{3}}\}])^{\exp\{Z_{i}\beta_{z}+m_{i}(t)\gamma\}}$ with $(\alpha_{3}, \lambda_{3}, \beta_{z}, \gamma, \beta_{30}, \beta_{31}) = (4, 0.02, 0.5, 0.8, 3, 2)$, where $p = 0.6$, $Z_{i}$ is a time-fixed covariate following Bernoulli(0.5), and $\gamma$ represents the effect of $Z_{i}(t)$. The type 2 failure times and censoring times were generated following the same way as specified before. 
 We chose a prediction window of $w = 0.4$ and sample size of $n = 5,000$ to calculate the 3-fold cross-validated time-dependent O/E ratio, Brier score, and AUC for the landmark PSH supermodel and compared it with the landmark Cox supermodel. To fit the supermodels, we selected landmark time points from 0 to 4 with a step of 0.1, and used the same quadratic basis function $f(s)$ and $g(s)$ for $Z_{i}(t)$ because that only the effect of $Z_{i}(t)$ depends on landmark $s$ in the Wald test. Table~\ref{tab:BSAUC} provides averaged estimates of the O/E ratio, Brier score, and AUC with the corresponding empirical standard errors from $1,000$ simulated data sets. The landmark PSH supermodel showed a good predictive performance in terms of calibration, predictive accuracy, and discrimination.

%******************************************
%    Application
%******************************************
%\clearpage
\section{Application \label{sec: application}}

To illustrate the application of our proposed landmark PSH supermodel in prediction of the conditional cumulative incidence probabilities for a moving (or dynamic) time interval with a fixed width, we used the data from a multicenter phase III clinical trial for breast cancer patients with estrogen receptor positive and historically nodes negative (Fisher et al.\ 1997). In this trial, 2,363 patients were randomly assigned to receive one of the following three regimens: tamoxifen 10mg daily for 5 years, tamoxifen 10mg daily for 5 years plus metrotrexate (M) and Fluorouracil (F), and tamoxifen 10mg daily for 5 years plus M, F, and cyclophosphamide (C); denoted TAM, TAM+MF, and TAM+CMF, respectively for simplicity. The median follow-up time was 11.2 years.

Among the 2,272 clinically eligible patients who were followed, 241 developed distant metastasis, 127 died due to other causes before distant metastasis could occur, and the remaining 1,904 were censored due to withdrawal of the study, lost to follow up, or event-free on or before the analysis cutoff date. In analyses, the censoring cases were treated as random noninformative censoring where the justification is described in Web Appendix C. For early stage breast cancer patients after surgery, development of locoregional recurrence (LRR) is an important prognostic clinical event affecting the risk of distant metastasis. In this data, $15.1\%$ of the patients developed LRR before progressing to distant metastasis; but only $6.6\%$ patients experienced LRR before death. Our main interest in this application is to dynamically predict the risk of distant metastasis within the subsequent 3-years for a breast cancer patient, based on her LRR status measured during follow-up and other prognostic covariates measured at baseline, including the treatment (TAM, TAM+MF, or TAM+CMF), surgery type (lumpectomy plus radiation therapy [L+XRT] vs. mastectomy), age at the study entry ($< 50$ vs. $\geq 50$ years old), clinical tumor size ($\leq 2$ vs. $> 2$ cm), and tumor grade (well, moderate, and poor). We also compared the dynamic 3-year fixed width probabilities of distant metastasis and death based on a patient's LRR history.

Web Figure 2(a) shows that the estimated CIFs for both distant metastasis and death in which the maximum event time around 13 years. Web Figure 2(b) depicts the estimated CIF of LRR. There were only a few random censoring events occurred during the first 10 years of follow-up (figure not shown). Thus, we chose the range of landmark points from 0 to 10 years and prediction window with a fixed width of 3-years. To fit a landmark PSH supermodel to this dataset, we took 51 equally spaced landmark points $s$ ($0 \leq s \leq 10$) and set the basis functions for $\bfbeta_{LM}(s)$ and $\bfgamma(s)$ as $\bfbeta_{LM}(s)=\theta_{1}+\theta_{2}s+\theta_{3}s^2$ and $\bfgamma(s)=\eta_{1}s+\eta_{2}s^2$, respectively. The frequencies of distant metastasis and death in each of the landmark sub-dataset for $s = 0, 1, \ldots , 10$ years are shown in Web Figure 2(c). 

%%%****** Figure 3: Basic description of B20 data ****************
%\begin{figure}[ht]
%\centering
%\includegraphics[height = 3in, width=3.5in]{Figure3.eps}
%\caption{(a) Nonparametric estimate of the cumulative incidence of distant metastasis and death; (b) Nonparametric estimate of the cumulative incidence of locoregional recurrence; (c) Frequencies of distant metastasis and death for each of the specified landmark data sets.}
%\label{fig:describe}
%\end{figure}

We began our analysis using the backward selection procedure to select those covariates of which  the effects were dependent on the landmark points. We tested the landmark-covariate interactions for each covariate via the Wald test based on the robust covariance matrix of the estimated coefficients from the landmark PSH supermodel. We found that for distant metastasis, the effects of TAM+MR and age were significantly dependent on the landmark points, whereas for death, only the effect of LRR status was dependent on the landmark points. The estimated coefficient and the corresponding robust standard error for a given prognostic factor are summarized in Table~\ref{tab: PBLM-PSHresult}. Multivariate Wald tests for the baseline parameters $(\eta_{1}, \eta_{2})$ were significant for both distant metastasis and death, indicating that the baseline subdistribution hazard also depended on the choice of landmark points. 

Given different LRR status ( including no LRR developed over the course of study, with the first LRR occurred at 3, 5, and 7 years) and different treatment regimens (TAM, TAM+MF, or TAM+CMF), Figure 3 %~\ref{fig: predCIF} 
depicts the predicted dynamic 3-year fixed width cumulative incidences of distant metastasis and death with the associated bootstrap $95\%$ confidence intervals for patients younger than 50 years old with poorly differentiated tumors, tumor size larger than 2cm and treated by L+XRT. If a patient did not develop LRR, the risk of having distant metastasis within the subsequent 3 years is very close to the risk of death in any treatment group. However, if the patient occurred LRR, she would have a much higher risk of developing distant metastasis compared to the risk of death, especially for TAM+MF treatment group. Similar trends were obtained for patients if they experience the first LRR at 3, 5, and 7 years, respectively.

We compared the risk prediction performance of our proposed landmark PSH supermodel against the landmark Cox supermodel. The fitted landmark Cox supermodel is summarized in Web Table 2. In addition to measure the time-dependent O/E ratio, Brier score, and AUC at a selected set of landmark time points with bootstrap standard errors with $B=500$, we also calculated the squared bias (Bias2) for both models. As results shown in Table~\ref{tab:AppBSAUC}, the landmark PSH supermodel is well-calibrated compared to the landmark Cox supermodel which tends to underestimate the risk of event with larger O/E ratios. Although two models provide similar Brier score results, the Cox supermodel has notably larger squared bias in predictive accuracy, which indicates that the Brier score is dominated by the Bernoulli variation of the outcome not by the squared bias in this application data, so leads to a limited variation in Brier score between two models. Because of the two models have the similar model form (Table~\ref{tab: PBLM-PSHresult} and Web Table 2), the difference of AUCs are limited in this application example. 

To demonstrate an additional value that dynamic prediction provides, we compared the performance of the landmark PSH supermodel that dynamically incorporated LRR status at each landmark time to the PSH model that simply used LRR status as a predictor to predict the CIFs at the horizon time. Results are summarized in Web Table 3. Apparently, the landmark PSH supermodel outperformed the PSH model in terms of both calibration and predictive accuracy.

%\clearpage
\section{Discussion \label{sec: discussion}}

In this study, we developed dynamic prediction models for data containing competing risks by extending the landmark approach to the Fine-Gray model. The resulting landmark PSH model and the supermodel can be used to directly predict the dynamic cumulative incidences for the occurrence of a specific event within a given prediction window of a fixed width through incorporating all available information up to the landmark time under the condition that the patient has not failed at the landmark time.

Our proposed models have several advantageous features over the currently available methods in predicting the dynamic cumulative inferences. First, our landmark PSH models can provide accurate estimates even when covariate effects are potentially time-varying and violate the PSH assumption whereas the model developed by Cortese et al.\ (2013) will lead to biased estimate under violations of the PSH assumption even after incorporating the landmark method into the Fine-Gray model. Second, unlike the Fine-Gray model which does not allow the use of internal time-dependent covariate in prediction of the CIF (Kalbfleisch and Prentice 2002, Beyersmann and Schumacher 2008), the landmark PSH models can incorporate both internal and external time-dependent information through a simple model form that bypasses modeling the covariate changing process and the time-to-event outcome process. The landmark method provides a simpler and explicit form of estimates and it is much easier to incorporate intermediate events and/or time-dependent covariates as compared with the multistate model and joint models that are more complicated and prone to overfitting. Furthermore, the proposed landmark PSH supermodel is more straightforward and simpler to implement using the existing standard software packages. In contrast with the landmark supermodel based on the cause-specific hazards (Nicolaie et al.\ 2013a), our landmark PSH supermodel predicts the conditional CIFs only in one step and provides a direct interpretation of the landmark-specific effects on the predictive probabilities. Compared with the landmark supermodel based on the pseudo-observations (Nicolaie et al.\ 2013b), our supermodel also shows simplicity in computation whereas the GEE-based method used in pseudo-observations would have convergence issues for large sample size especially when dealing with many landmark points of interest.

We evaluated the prediction performance of our proposed models through simulations and compared to other existing methods. We determined how closely the estimated conditional CIFs would approximate the true probabilities for our models, which is the first of its kind done in this area of study. We also evaluated the landmark supermodel in competing risks data from the perspectives of calibration, predictive accuracy, and discrimination capabilities by utilizing the time-dependent O/E ratios, Brier scores and AUCs. The simulation results showed that our models performed well in prediction after incorporating time-fixed and time-dependent covariates even when the PSH assumption was violated. 

In the landmark PSH supermodel, we assume that the effect of $s$ on baseline hazards $\lambda_{10}(t|s)$ was an additive effect. As discussed in van Houwelingen (2007), this assumption will hold if the follow-up is not too long or the effect of covariates is not too large. If we choose an optimal width $w$ for the prediction window and a suitable range $[s_{0}, s_{L}]$ for the landmark time points, the landmark PSH supermodel can provide a correct approximation for the conditional CIF at time $s+w$ for any $s\in [s_{0}, s_{L}]$. An alternative method is to fit a stratified landmark PSH supermodel by applying a Fine-Gray model with the landmark-covariate interactions $\bm{Z}(s)*f_{j}(s)$ to the stacked dataset with stratification on $s$, which estimates the baseline conditional subdistribution hazard for each $s$. 

Note that in the presence of time-dependent covariates, if the covariate values vary too often and too much over time, the estimated covariate effects could be attenuated in the proposed models and subsequently lead to biased estimation of the conditional CIF. This is also shown in our simulation result for data with a time-dependent covariate. To adjust for this issue, which was also discussed by van Houwelingen and Putter (2012), we will explore a set of suitable landmark time points and use an additional monotonically decreasing function to model the attenuation process of the covariate effects as future work.

\clearpage
%============ Figures ========= %
%******* Figure 1: Predicted Conditional CIF ****************
\begin{figure}[!p]
  \centering
  \includegraphics[height=3in,width=3.5in]{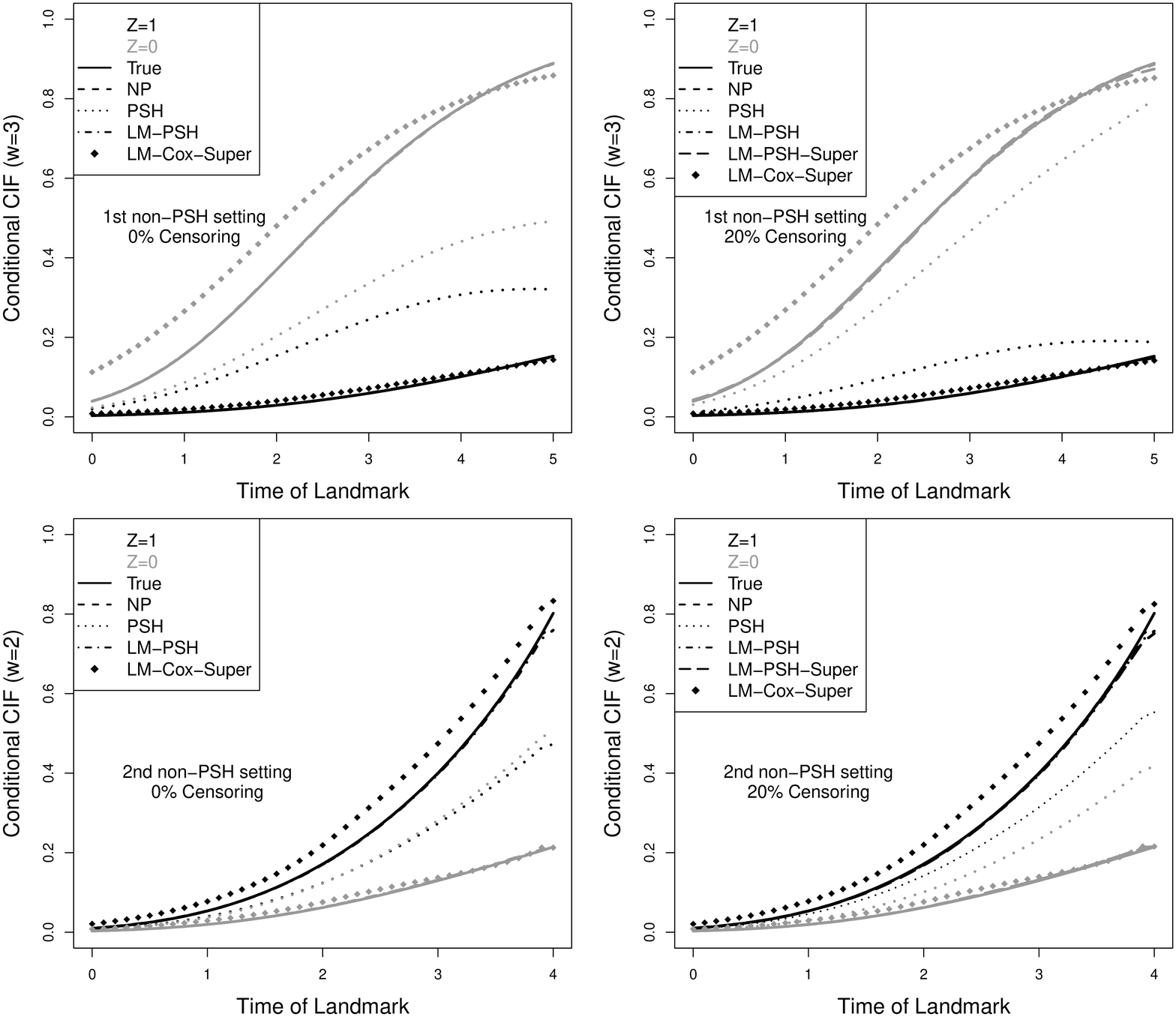} 
\caption{Predicted conditional cumulative incidences over $w$ years (averaged over 1000 simulations) at landmark points. The 1st non-PSH setting with $(\alpha_{1}, \lambda_{1}, \beta_{1}) = (3.2, 0.18, -0.81)$ and the 2nd non-PSH setting with $(\alpha_{2}, \lambda_{2}, \beta_{21},\beta_{22}) = (3.2, 0.12, 0.8, 0.3)$. $Z$ is Bernoulli(0.5) variant; and $p = 0.3$, which produced about $30 \%$ main events at $Z_{i}=0$ when there was no censoring. Black curves: the $Z=1$ group; gray curves: the $Z=0$ group. True: underlying true conditional CIFs. NP: nonparametric method; PSH: the standard PSH model; LM-PSH; the landmark PSH model; LM-PSH-Super: the landmark PSH supermodel; LM-Cox-Super: the landmark Cox supermodel, in which the y-axis is the conditional failure probabilities for main event.}
  \label{fig: predictedcondCIF}
\end{figure}

%****** Figure 2: CVBS ****************
\begin{figure}[!p]
  \centering
  \includegraphics[height=3in,width=3.5in]{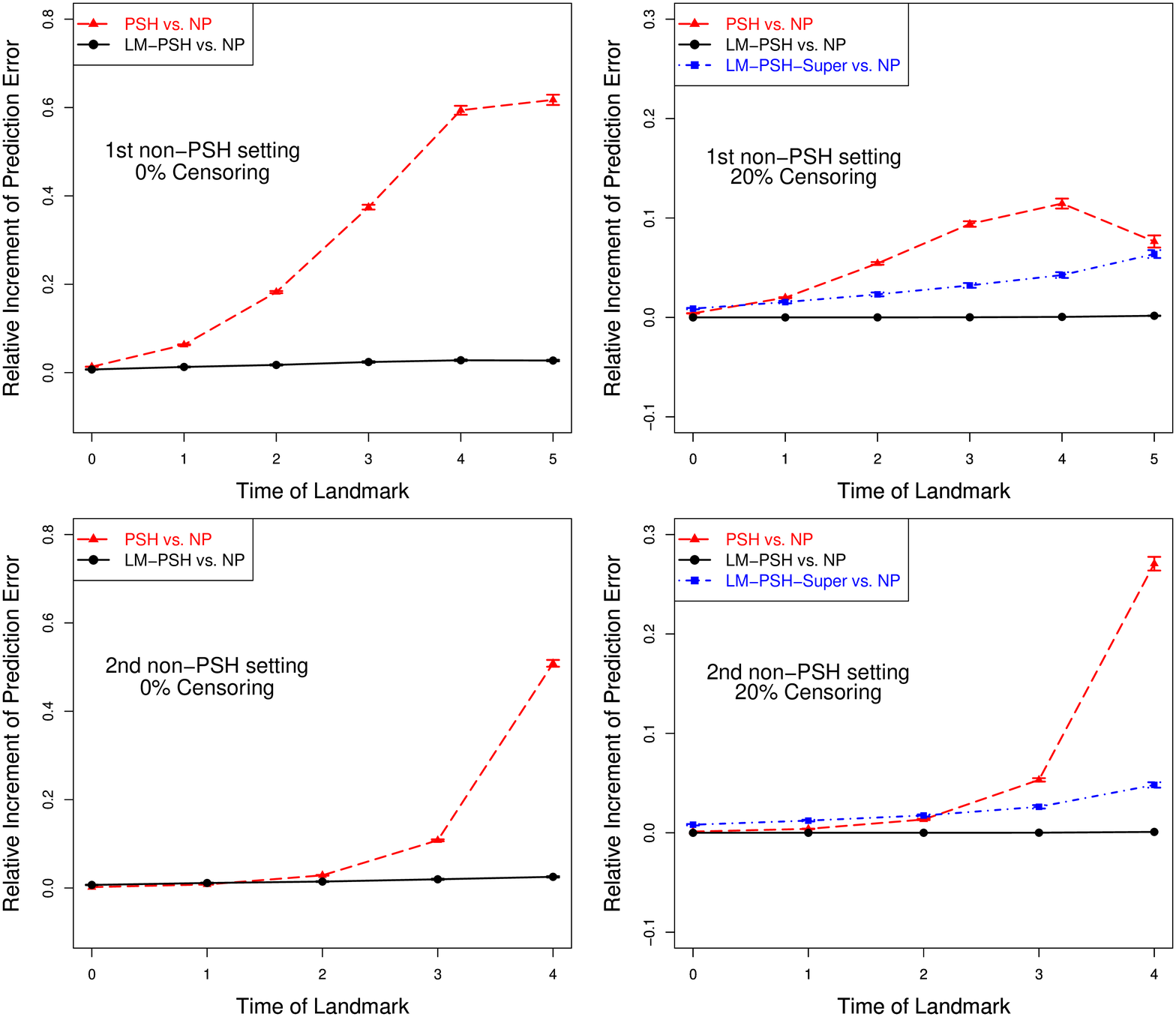} 
  \caption{Relative increment of prediction errors (and their standard deviation) at landmark points.  The prediction errors were cross-validated (3-fold) estimates for the time-dependent Brier scores, where $w=3$ for the 1st non-PSH setting with $(\alpha_{1}, \lambda_{1}, \beta_{1}) = (3.2, 0.18, -0.81)$ and $w=2$ for the 2nd non-PSH setting with $(\alpha_{2}, \lambda_{2}, \beta_{21},\beta_{22}) = (3.2, 0.12, 0.8, 0.3)$. $Z$ is Bernoulli(0.5) variant; and $p = 0.3$, which produced about $30 \%$ main events at $Z_{i}=0$ when there was no censoring. NP: nonparametric method; PSH: the standard PSH model;  LM-PSH; the landmark PSH model; LM-PSH-Super: the landmark PSH supermodel.}
  \label{fig: cvbs}
\end{figure}

%****** Figure 4: Plot predicted curves ****************
\begin{figure}[!p]
  \centering
    \includegraphics[height=7in,width=6.5in]{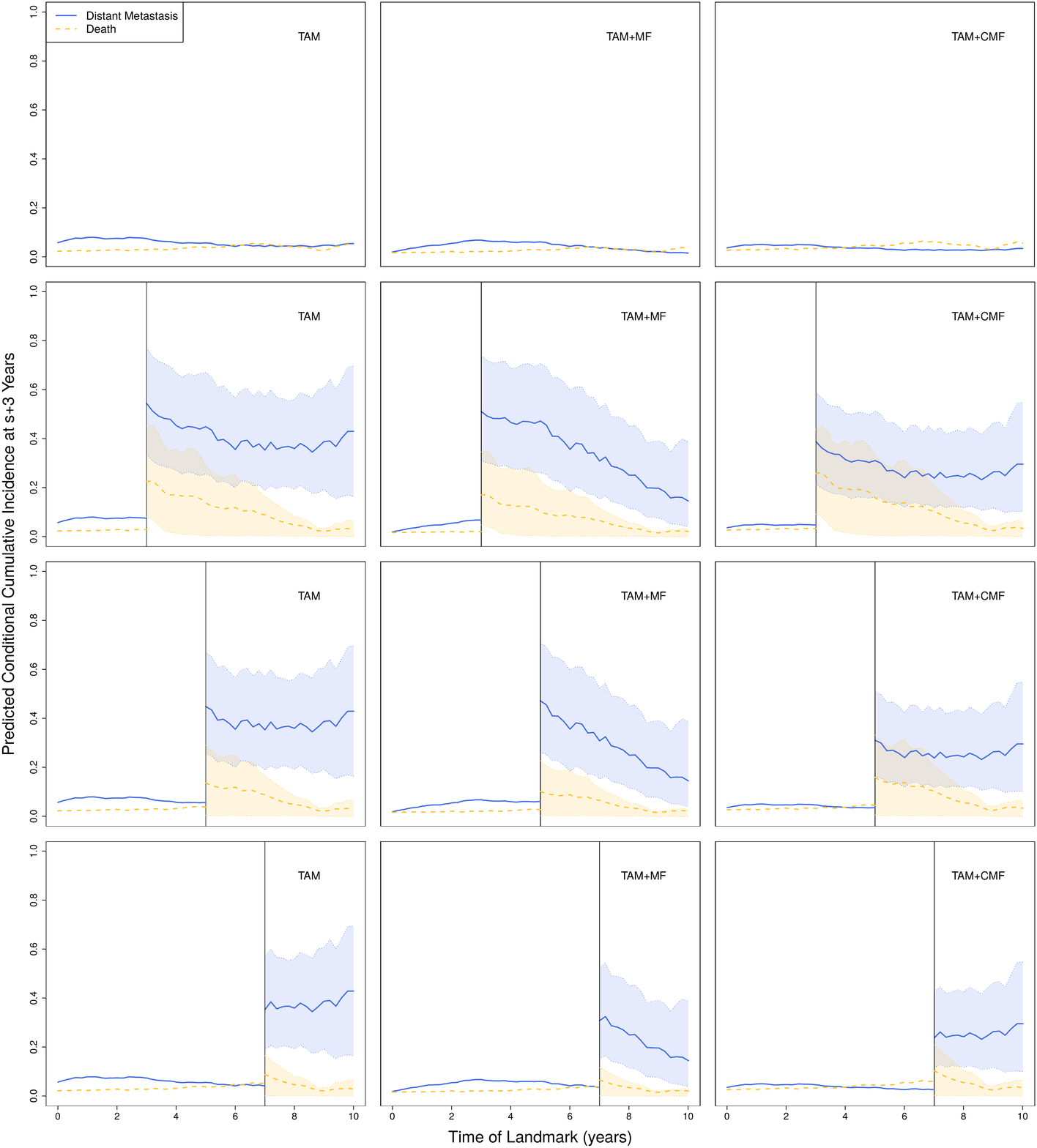} 
  \caption{The predicted 3-year fixed width conditional cumulative incidences of distant metastasis (blue solid lines) and death (yellow dashed lines) with the associated bootstrap $95\%$ confidence intervals (shaded areas) for different landmark time points, for patients younger than 50-years old with poor tumor grade, tumor size larger than 2cm and treated with lumpectomy in each of the treatment groups and with the first locoregional recurrence occurred at  none, 3 years, 5 years and 7 years. }
  \label{fig: predCIF}
\end{figure}

%\clearpage
%============ Tables ========== %
\begin{table}[!p]
\centering
\caption{Cross-validated (3-fold) estimates of the time-dependent O/E ratio of observed number of events (O) to expected number of events (E), Brier score and AUC with the corresponding standard errors (se) for the landmark PSH supermodel (LM-PSH-Super) and the landmark Cox supermodel (LM-Cox-Super) in the 2nd non-PSH setting with time-fixed and time-dependent covariates. All entries are multiplied by 100.}
\label{tab:BSAUC}
\begin{tabular}{@{}cccc@{}}
   \hline
     & \multicolumn{3}{c}{LM-PSH-Super}  \\ 
     Landmark &  O/E (se) & Brier score (se) & AUC (se) \\
    \hline
     2.4 &  103.001 (15.356)  & 4.237 (11.421)  & 67.285 (5.805)  \\
     2.6 &  101.166 (14.050)  & 5.511 (11.836)  &  68.113 (4.545)  \\ 
     2.8 &  99.082 (12.702)   & 6.906 (11.135)  &  70.396 (3.749)  \\
     3.0 &  93.396 (12.686)   & 9.799 (11.981)   & 74.628 (3.966)  \\
     3.2 &  95.898 (14.165)   &  13.664 (12.298)  & 74.143 (4.253)  \\ 
     3.4 &  96.497 (13.877)   &  16.807 (11.093)  & 73.681 (4.343)   \\ 
     3.6 &  96.486 (13.257)   &  19.360 (9.965)  & 73.741 (4.618) \\                                                                                                                       
        \hline
        & \multicolumn{3}{c}{LM-Cox-Super} \\
        Landmark &  O/E (se) & Brier score (se) & AUC (se) \\
        \hline
        2.4 &   70.687(44.813)   & 37.405 (36.391) & 60.853 (8.206) \\
             2.6 &  65.017 (45.379)   & 41.575 (37.761) & 60.820 (8.270) \\ 
             2.8 &  61.776 (44.256)   & 43.415 (38.241) & 61.910 (9.160)\\
             3.0 &  56.393 (41.745)   & 47.426 (37.717)  & 63.376 (11.370) \\
             3.2 &  59.426 (42.258)   & 48.259 (34.891)  & 62.858 (11.656) \\ 
             3.4 &  65.245 (39.696)   & 44.926 (30.568)  & 63.480 (11.737) \\ 
             3.6 &  66.728 (34.844)   & 43.217 (25.862)  & 63.185 (11.674)\\                                                                                                                       
                \hline
\end{tabular}
\end{table}

%************* Table 2: PBLM-PSH model regression results *********************
\begin{table}[!p]
\caption{Estimated regression parameters of the landmark PSH supermodel for distant metastasis and death. $s$ is landmark time; $\eta_{1}$ and $\eta_{2}$ are baseline parameters defined in Section~\ref{subsec: lmfgsuper}.}
\label{tab: PBLM-PSHresult}
\begin{tabular}{@{}lcrcrc@{}}
    \hline
       &      &  \multicolumn{2}{c}{Distant metastasis}     &  \multicolumn{2}{c}{ Death}     \\
\hspace{.6in} Covariate   &      &  $  \hat{\bfbeta}$   &  se($\hat{\bfbeta}$)  &  $\hat{\bfbeta}$   &  se($\hat{\bfbeta}$)    \\
    \hline
    Treatment &  &  &  &  & \\
\hspace{.2in} {\small{TAM + MF vs. TAM}}& Constant  &  -1.091   &  0.331  &  -0.312 &  0.238   \\
  & $s$  &  0.482  & 0.160   &   &     \\
  & $s^2$  & -0.050   & 0.017   &   &     \\
\hspace{.2in} {\small TAM + CMF vs. TAM} & Constant  &  -0.470  & 0.170   & 0.160  & 0.219    \\
  Age &  &  &  &  &   \\
 \hspace{.2in} {\small $\geq 50$ vs. $< 50$} & Constant  &  -0.540  & 0.257  & 1.057   & 0.218    \\
  & $s$  &  0.407  & 0.135   &   &     \\
  & $s^2$  & -0.035   & 0.015   &   &     \\
Surgery type&  &  &  &  & \\
\hspace{.2in} {\small Mastectomy vs. L + XRT} & Constant  &  0.424  & 0.144   & -0.115  &  0.189   \\
Clinical tumor size &  &  &  &  &   \\
\hspace{.2in} {\small $>2$ vs. $\leq 2$ cm} & Constant  & 0.284   & 0.141   & 0.334  &  0.196   \\
Tumor grade  &  &  &  &  & \\
\hspace{.2in} {\small Moderate vs. Well}  & Constant  & 0.221   &  0.179  &  0.104 & 0.221    \\
\hspace{.2in} {\small Poor vs. Well}  & Constant  & 0.752   & 0.189   & 0.124  &  0.258   \\
Localregional recurrence status   &  &  &  &  & \\
 & Constant  & 2.315   & 0.237   & 3.575  &  1.230   \\
   & $s$  &    &    & -0.492  &  0.582   \\
   & $s^2$  &    &    & 0.008  &  0.056   \\
Baseline parameters  &  &  &  &  & \\
\hspace{.2in} {\small$\eta_{1}$}& $s$  & -0.400   &  0.088   & -0.001  & 0.014    \\
\hspace{.2in} {\small  $\eta_{2}$}& $s^2$  &  0.035  & 0.009   & 0.002  &  0.001   \\
    \hline
\end{tabular}
\end{table}

%************* Table 3: App BS and AUC *********************
\begin{table}[!p]
\caption{Estimates of the time-dependent O/E ratio of the observed number of events (O) to the expected number of events (E), squared bias (Bias2), Brier score and AUC (with standard errors from bootstrap with B=500) at a selected set of landmark time points for distant metastasis and death. LM-PSH-Super: the landmark PSH supermodel; LM-Cox-Super: the landmark Cox supermodel. All entries are multiplied by 100.}
\label{tab:AppBSAUC}
\begin{tabular}{@{}lccccc@{}}
    \hline
    & Landmark & \multicolumn{4}{c}{LM-PSH-Super} \\ 
    & (years) &  O/E (se) & Bias2 (se) & Brier score (se) & AUC (se)  \\
    \hline
                        & 1 &  102.055 (10.042)  & 0.00008 (0.003) & 4.237 (0.395)  & 66.174 (2.943)  \\
     Distant        & 3 &  98.609 (9.938)      & 0.00003 (0.002) & 3.854 (0.370)  &  68.260 (3.156)  \\ 
     Metastasis   & 5 &  99.134 (12.911)    & 0.000007 (0.002) & 2.856 (0.364)  &  74.688 (3.266)  \\
                        & 7 &  99.771 (15.288)   & 0.0000003 (0.002)& 2.070 (0.308)   & 67.821 (4.718)  \\
    \hline
                        & 1 &  100.754 (19.187)  & 0.0000008 (0.0008) & 1.206 (0.224) & 62.136 (6.003)  \\
        Death       & 3 &  97.965 (16.038)    & 0.000009 (0.0008) & 1.439 (0.228)  & 76.705 (4.398)  \\
                        & 5 &  100.202 (16.506)  & 0.0000002 (0.002) & 1.913 (0.308)  & 64.165 (4.530)  \\
                        & 7 &  100.208 (14.454)  & 0.0000003 (0.002) & 2.553 (0.357)  & 63.816 (4.282)  \\
    \hline
    & Landmark & \multicolumn{4}{c}{LM-Cox-Super} \\ 
    & (years) &  O/E (se) & Bias2 (se) & Brier score (se) & AUC (se)  \\
    \hline
                        & 1 &  101.099 (9.949)  & 0.00002 (0.003) & 4.265 (0.397)  & 66.175 (2.953)  \\
     Distant        & 3 &  124.565 (12.581) & 0.007 (0.008)  & 3.881 (0.382)  &   68.373 (3.151) \\ 
     Metastasis   & 5 &  111.843 (14.550)   & 0.001 (0.004)  & 2.885 (0.371)  & 74.639 (3.264)   \\
                        & 7 &   120.795 (18.517)  & 0.001 (0.004) & 2.097 (0.316)    & 67.917 (4.692)  \\
    \hline
                        & 1 & 103.829 (19.861)   & 0.00002 (0.0009) & 1.241 (0.228) & 62.291 (6.009)   \\
        Death       & 3 & 110.630 (18.211)   & 0.0002 (0.001) & 1.478 (0.235)  &  76.497 (4.391) \\
                        & 5 & 106.203 (17.515)   & 0.0001 (0.002) & 1.947 (0.314)  & 64.295  (4.539)\\
                        & 7 & 109.095 (15.754)   & 0.0005 (0.003) & 2.585 (0.363)  & 64.924 (4.217)\\
    \hline
\end{tabular}
\end{table}

\end{document}